\documentclass[aps,preprint]{revtex4}
\usepackage{graphicx}
\usepackage{amssymb}
\usepackage{amsmath}
\usepackage{color}
\usepackage{enumerate}
\usepackage{bm}
\begin{document}

\title{Controlling degeneracy and magnetization switching in an  artificial spin ice system of peanut-shaped nanomagnets}

\author{Avinash Chaurasiya}
%\email{co-first authors}
\altaffiliation{These authors contributed equally to this work}
\affiliation{Natural Sciences and Science Education, NIE, Nanyang Technological University, 637616, Singapore}

\author{Manish Anand}
%\email{co-first authors}
\altaffiliation{These authors contributed equally to this work}
\affiliation{Department of Physics, Bihar National College, Patna University, Patna-800004, India.}

\author{Rajdeep Singh Rawat}
\email{rajdeep.rawat@nie.edu.sg}
\affiliation{Natural Sciences and Science Education, NIE, Nanyang Technological University, 637616, Singapore}

\date{\today}

\begin{abstract}
Using extensive numerical simulations, we probe the magnetization switching in a two-dimensional artificial spin ice (ASI) system consisting of peanut-shaped nanomagnets. We also investigated the effect of external magnetic field on the degeneracy of the magnetic states in such a system. The switching field is found to be one order smaller in the proposed ASI system with peanut-shaped nanomagnets as compared to the conventionally used highly-anisotropic nanoisland such as elliptically shaped nanomagnets. The metastable 2-in/2-out (Type II) magnetic state is robust at the remanence. We are also able to access the other possible microstate corresponding to Type II magnetic configurations by carefully varying the external magnetic field. It implies that one can control the degeneracy of the magnetic state by an application of suitable magnetic field. Interestingly, the magnetic charge neutrality breaks due to the defects induced by removing nanomagnets. In such a case, the system also appears to have 1-out/3-in or 3-out/1-in (Type III) spin state, reminiscent of magnetic monopole. We believe that our study is highly desirable in the context of developing the next-generation spintronics-based devices for future technologies.
\end{abstract}
\maketitle

\section{Introduction}
Artificial spin ice (ASI) systems have received significant attention in recent years due to their rich physics and various technological applications~\cite{nisoli2010,lendinez2019,skjaervo2020,kapaklis2014,nisoli2013,zhang2019}. They are ideal systems to investigate numerous intriguing physical phenomena such as frustration in magnetism, emergent magnetic monopoles, magnetization dynamics, phase transition, etc.~\cite{wang2006,morgan2011,jungfleisch2017,gilbert2014,sen2015}. On the other hand, they have potential applications in spintronics,  re-programmable magnonic crystals, data storage, logic gates, etc.~\cite{iacocca2016,mamica2012,nikitin2015,arava2019,wang2016}.

%one of the most promising as-pects of ASI system is their high tunability of the array geometry and nanomagnetparameters which has become possible due to the availability of the modern lithog-raphy techniques.  ASI can be considered as model systems to study the physicsof frustration in a controlled way thereby allowing exploring new avenues which isnot always possible in natural materials.

ASI systems are lithographically patterned arrangements of interacting nanomagnets with strong shape anisotropy. They are generally arranged in a square, hexagonal, honeycomb lattices, etc., and the individual nanomagnets behave as Ising-like macro-spins because of high anisotropy~\cite{kapaklis2012,shen2012,qi2008,ladak2010}. One of the most important aspects of ASI structures is the excitation of emergent magnetic monopoles and their controllability, which is generally characterized using a dumbbell model ~\cite{nascimento2012}, as shown in Fig.~(\ref{figure1}) and Fig.~(\ref{figure2}). In this context, various works have extensively investigated how to control the monopoles and the magnetization switching of constituents magnetic  nanoislands ~\cite{hugli2012,castelnovo2008,arava2020,mengotti2011,loreto2014,bramwell2012,ladak2011,keswani2018}.  For example, using experiments and Monte Carlo simulation, Arava {\it et al.} demonstrated the control of emergent magnetic monopole current in ASI system~\cite{arava2020}. Mengotti {\it et al.} reported real-space observations of emergent monopoles in two-dimensional artificial kagome spin ice using synchrotron X-ray photoemission electron microscopy~\cite{mengotti2011}. Loreto {\it et al.} also studied the physics of emergent magnetic monopole in the presence of an external magnetic field~\cite{loreto2014}. Bramwell analyzed the spatial and temporal correlations of magnetic monopoles in spin ice using the generalized longitudinal susceptibility~\cite{bramwell2012}. Ladak {\it et al.} investigated the formation of monopole defects in a cobalt honeycomb ASI system using micromagnetic simulations and experiments~\cite{ladak2011}. They observed that monopole defects of opposite sign are created at the boundaries of the lattice which move in opposing directions. In a recent study, Keshwani {\it et al.} investigated the emergence of magnetic monopoles in an ASI system consisting of elliptical-shaped nanomagnets. They reported that such magnetic configurations could be created and controlled by manipulating defects and external magnetic fields~\cite{keswani2018}.

These works suggest that the study of the emergence of magnetic monopoles in the ASI systems is of vital importance. It is also equally important to investigate the detailed mechanism to control the magnetization switching of constituents nanomagnets in such a versatile system. Some recent works suggest that one can also use the ASI system to control the magnetization switching through the piezoelectric strain-mediated magneto-electric (ME) effect~\cite{chavez2018,kaffash2021,frotanpour2020,chaurasiya2020nickel}. The strain-based magnetization switching in the ASI system has high potential in the next generation spintronics devices ~\cite{wolf2001}. The magneto-electric effect driven nanoisland (nanoscale magnetic elements) can store bit information without any significant power dissipation, thus offering unprecedented power efficiency~\cite{wolf2001}.
%The magnetoelectric driven nanomagnets 
%{\color{black} ME driven nanomagnets (nanoscale magnetic elements) can store bit information without any standby power dissipation, offering unprecedented power efficiency, and thus have high potential for implementation in the next generation of spintronic devices~\cite{wolf2001}}.
However, it is difficult to obtain complete magnetization reversal by strain alone because of its uniaxial character~\cite{hu2009,wang2014}. Therefore, the possible usage of such ASI structure is limited as one generally uses highly anisotropic nanoisland such as elliptical-shaped nanomagnet [please see Fig.~\ref{figure3}(a)], whose easy and hard axes are perpendicularly aligned~\cite{keswani2020,keswani2019,buzzi2013}. To overcome this shortcoming, we need to design an ASI system of nanomagnets that have an angle between the easy and hard axes of less than $90^\circ$. {\color{black} However, the goal of achieving a decrease in the switching magnetic field leads to the reduction in the symmetry of the ASI system. Interestingly it is found that instead of reduction in symmetry,    we have achieved all possible switching states for the peanut shaped ASI system similar to the elliptical shaped ASI system.} Fig.~\ref{figure3}(a-b) shows the orientation of the {\color{black} hard} and the {\color{black} easy} axis of elliptical and peanut-shaped nanomagnet, respectively. Here, the elliptical nanomagnet shows the {\color{black} hard} and {\color{black} easy} axis along the $x$ and $y$-axis, respectively. The demagnetization energy of the ellipse along the {\color{black} easy} axis is larger in comparison to {\color{black} hard} axis which results in the direction of the easy axis of magnetization along the longer axis of ellipse. By deforming the ellipse, we can construct the peanut-shaped nanomagnet (see Fig.~\ref{figure3}(b)). The designed peanut shape also has easy axis of magnetization along the longer axis. The orientation of the easy axis of magnetization of the designed system is calculated by finding the variation of the demagnetization energy with the angle of rotation of constant applied magnetic field, shown in Fig.~\ref{figure3}(c). It is clear that the minima of the demagnetization energy is at about $25^\circ$ and the maxima is at $85^\circ$, as shown in Fig.~\ref{figure3}(c). Therefore, the angle between the {\color{black} hard} and the {\color{black} easy} axis is about $60^\circ$ for peanut-shaped nanomagnet (less than from conventional elliptical-shaped nanomagnet).
Thus motivated, we propose an engineered ASI structure consisting of peanut-shaped nanomagnet in the present work [see Fig.~(\ref{figure4})].
%the strain-mediated electrically driven magnetization switching process is not deterministic~\cite{}. 
%strain is a uniaxial effect and, unlike directional magnetic field or spin-polarized current, cannot induce a full $180^\circ$ reorientation of the magnetization vector~\cite{}. 
%In ASI system, one generally uses highly anistropic nanoisland whose easy and hard axis are aligned at $90^\circ$ apart. Therefore, the possible usage of such ASI is limited as the strain-mediated electrically driven magnetization switching process is not deterministic~\cite{}. 
As the angle between the easy and hard axis is less than $90^\circ$ in a peanut-shaped nanomagnet, one can achieve complete
magnetization reorientation controlled by electric-field-induced strain~\cite{cui2017}. To probe the physics of magnetization reversal and various aspects of the ASI system, we have performed extensive micromagnetic simulations and investigated the switching behaviour of an ASI structure consisting of peanut-shaped nanomagnets arranged in the square arrangement as depicted in Fig.~\ref{figure4}(a). We have also probed the effect of defects (missing nanomagnet) [see Fig.~\ref{figure4}(b)-(d)] on the variation of coupling between the individual nanomagnets, which leads to a change in the magnetic behaviour. Our study suggests that we can control the formation of monopoles-like states and the degeneracy of the magnetic configurations by carefully controlling the roles of defects and the external magnetic field.

%The rest of paper is organized as follows. In Sec. II, we present the
%model and discuss the various energy terms and methods of simulations. The simulations results will be presented in Sec. III. Finally, we summarize and
%discuss the main results in Sec. IV. 

\section{Model}
To investigate the magnetization switching behaviour of an ASI structure,
it is essential to consider the required energy terms such as the exchange energy, the demagnetization energy, and the Zeeman energy (due to an external magnetic field). Considering all the above-mentioned energy terms, we can write the total energy of a system with volume $V$ as~\cite{salehi2017}
\begin{equation}
\begin{aligned}
E={} &\oint \bigg[A\{(\nabla m_x)^2+(\nabla m_y)^2+(\nabla m_z)^2\}-
\frac{1}{2}\mu^{}_o\vec{H}^{}_d\cdot\vec{M}-\mu^{}_o\vec{M}^{}\cdot\vec{H}
\bigg]dV
\label{eq1}
\end{aligned}
\end{equation}
Where $\nabla m_i$ are the magnetization gradients, $\mu^{}_o\vec{H}^{}_d$ is the demagnetization field, $\mu^{}_o\vec{H}$ is the applied external magnetic field, and $\vec{M}$ magnetization of the nanomagnets.
In Eq.~(\ref{eq1}), the first term on the right denotes the exchange energy having exchange constant $A$. The second term is the demagnetization energy of the nanomagnet while the last term represents the Zeeman energy of interaction with an external magnetic field $\mu^{}_o{H}$.  In the present work, magneto-crystalline anisotropy is neglected as the nanomagnet is assumed to have random polycrystalline orientation~\cite{wang2014full}. The corresponding demagnetization field $\mu_o \vec{H}^{}_d$ is given by the following expression~\cite{anand2016,miltat2007}
\begin{equation}
\mu^{}_o\vec{H}^{}_d=-\frac{1}{4\pi}\int_{V}\vec{\nabla}\cdot{ \vec{M}(r')\frac{\vec{r}-\vec{r'}}{|\vec{r}-\vec{r'}|^{3}}}d^3r+\frac{1}{4\pi}\int_{S}\hat{n}\cdot{ \vec{M}(r')\frac{\vec{r}-\vec{r'}}{|\vec{r}-\vec{r'}|^{3}}}d^2r    
\label{dipolar}  
\end{equation}
here $\mu^{}_o$ is the permeability of free space, $\hat{n}$ is the surface normal and $\vec{r}$ and $\vec{r'}$ are the position vectors of two different nanomagnets. 

The magnetization dynamics of a nanomagnet under the influence of the total or effective magnetic field is described by the Landau-Lifshitz-Gilbert (LLG) equation as~\cite{anand2016,arora2021}
\begin{equation}
\frac{d\vec{M}(t)}{dt}=-\gamma\vec{M}(t)\times\vec{H}^{}_{\mathrm {eff}}(t)-\frac{\alpha\gamma}{M^{}_s}\big[\vec{M}(t)\times(\vec{M}(t)\times\vec{H}^{}_{\mathrm {eff}}(t))\big]
\label{llg}
\end{equation}
Here $\vec{H}^{}_{\mathrm {eff}}$ is the effective magnetic field acting on the nanomagnet expressed as~\cite{anand2016}
\begin{equation}
\vec{H}^{}_{\mathrm{eff}}=-\frac{1}{\mu^{}_oV}\frac{\partial E(t)}{\partial \vec{M}^{}(t)}
\end{equation}
$\gamma$ is the gyromagnetic ratio, $M^{}_s$ is the saturation magnetization and $\alpha$ is the damping factor associated with internal dissipation in the magnet owing to the magnetization dynamics. 

%Defining stress, $\sigma=\epsilon Y$, where $Y$ is the Young’s modulus, $\epsilon$ is the strain,
%We used a finite difference approach by discretizing the magnetic samples %into cubic cells. The computations were performed using three dimensional solver of the Object Oriented Micro Magnetic Framework (OOMMF) free software from the National Institute of Standards and Technology

We have performed micromagnetic simulations to simulate the  magnetic moment interactions and magnetization switching dynamics in the array of peanut-shaped nanomagnet structure. For it, we have used Object Oriented MicroMagnetic Framework (OOMMF) software from the National Institute of Standards and Technology~\cite{donahue1999}. It is based on the finite-difference approach by discretizing the magnetic samples into cubic cells, which is extensively used to investigate the magnetization dynamics in such a system~\cite{donahue1999}.
%based micromagnetic simulation to simulate the  magnetic moment interactions and magnetization switching dynamics in the array of peanut-shaped structure of ferromagnet.
%We have used Object Oriented MicroMagnetic Framework (OOMMF) software (ref) in order to explore magnetic moment interactions and magnetization switching dynamics in the array of peanut-shaped structure. 
%It is based on the continuum theory of micromagnetics which is used to describe the magnetization process within ferromagnetic material. 
%To analyze the magnetization reversal process and time evaluation of magnetic moment in the peanut shaped structure, three dimensional micromagnetic simulations were performed using OOMMF. These OOMMF simulations perform time integration of the LLG equation, where the effective  energy includes the exchange, anisotropy, self-magnetostatic and external magnetic fields. 
The discretized cell size used in the simulations is 1000 nm $\times$1000 nm$\times$5 nm, implemented in the cartesian co-ordinate system. The parameters used for simulation are as follows: exchange constant $A =13\times10^{-12}$ Jm$^{-1}$, saturation magnetization $M^{}_s=8.60\times10^5$ $A\mathrm{m^{-1}}$, anisotropy constant (magnetocrystalline anisotropy) $K = 0$, damping coefficient $\alpha= 0.02$. These parameters correspond to permalloy, one of the ideal materials for such a study~\cite{gartside2018}. The magnetic switching of the islands are studied by sweeping the field within $\pm30$ mT, and it is applied along the $x$-axes of the system. The careful ramping up or slowing down of the magnetic field has been appropriately fine-tuned to capture the physics near the coercive field.  

\section{Simulation Results}
\begin{figure}[!htb]
	\centering\includegraphics[scale=0.10]{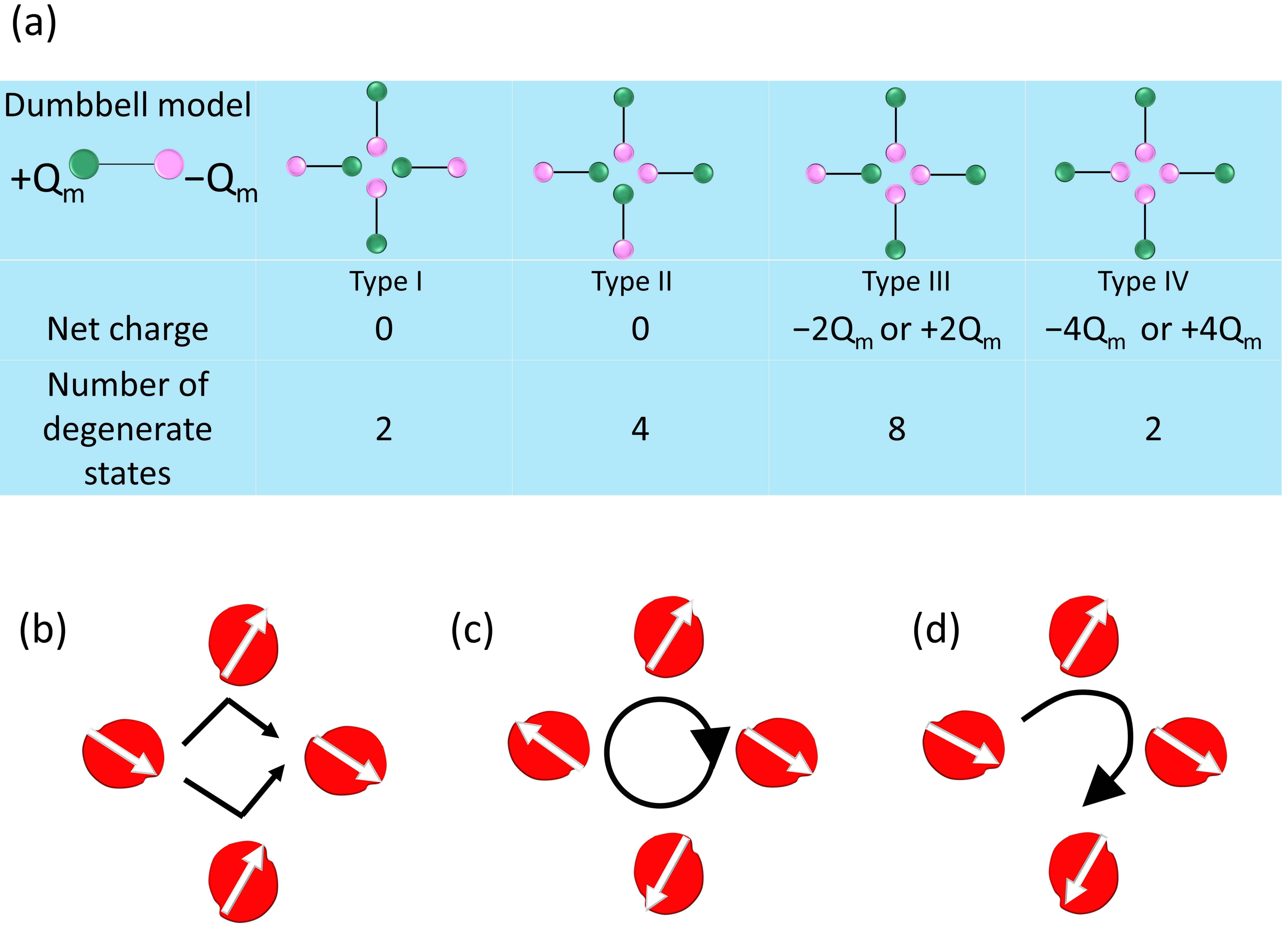}
	\caption{(a) Possible spin states at a vertex of square artificial spin ice system with peanut-shaped nanomagnet. The spin states at the vertex characterized by four different types with degeneracy of 2, 4, 8 and 2 respectively. The charge at the vertex defined by using the dumbbell model (the head with positive and tail with a negative charge).
		Possible spin states at the vertex (b) onion , (c) micro-vortex state (d) and horse-shoe~\cite{remhof2008} .}
	\label{figure1}
\end{figure}

We first probe the magnetization switching and study the remanent state in an isolated vertex with edges open (no edge-island) [see Fig.~\ref{figure4}(a)]  ASI system. Fig.~(\ref{figure5}), shows the magnetization as a function of the external magnetic field and corresponding spin state at various representative magnetic fields for open edge vertex. The variation of the demagnetization and the exchange energy as a function of external magnetic field has also been plotted to elucidate the role of these energies in switching the peanut-shaped nanomagnets of the under-lying system(see Fig.~\ref{figure5}(b)).
%at different. spin states corressponding to various magnetic field values. We found that the strength of demagnetization energy is one order higher in comparison to the exchange energy counterpart as the system is highly shape anisotropic in nature.
%is swept. We have also shown the 
It is obvious from Fig.~\ref{figure5}(a) that the saturation or anisotropy field $\mu^{}_oH^{}_K$ is about 30 mT, which corresponds to the anisotropy strength $K^{}_{shape}$ (shape anisotropy in the present case)$\approx 1.29\times 10^4$ Jm$^{-3}$. It is also clear from the shape of the field-dependent magnetization curve [Fig.~\ref{figure5}(a)] that the angle between the anisotropy axis (due to shape) and hard axis is less than $90^\circ$~\cite{carrey2011}, as the ratio of remanent and saturation magnetization is less than one. Interestingly, the first prominent magnetization switching occurs at the coercive field $\mu^{}_oH^{}_c=5$ mT, as evident from Fig.~\ref{figure5}(a). Keshwani {\it et al.} reported such occurrence of switching at 134 mT for an ASI structure consisting of elliptical nanomagnets ~\cite{keswani2018}. Therefore, it implies that it is easier to rotate the magnetization in the present system with peanut-shaped nanomagnets~\cite{keswani2018} [please see Fig.~S1 and Fig.~S2 in the Supplementary Information]. In an ASI system, it is essential to control the magnetization switching. Wang {\it et al.} reported such controlled switching of individual nanostructures by applying magnetic field pulses using tip of a magnetic force microscopes~\cite{wang2016}. In this context, the present ASI structure with peanut-shaped nanomagnets could be a better alternative than other ASI structures consisting of highly anisotropic nanomagnets as one needs a minimal magnetic field to control the magnetization switching. Remarkably, we also observe a 2-in/2-out (Type-II) magnetic state at $\mu_oH^{}_c=-5$ mT as shown in Fig.~\ref{figure5}c(i). Similarly, we observe one of the other possible configurations with 2-in/2-out (Type-II) state at 0 mT [Fig.~\ref{figure5}c(v)]. {\color{black} Importantly, as shown in Fig.~\ref{figure1}(a)  type II structure has four degenerate states and out of four possible state one of the state is shown in Fig.~\ref{figure1}(b). Here for a given system (vertex with open edges) at +5 mT field one of the possible state is achieved correspondingly the opposite and other degenerate state is obtained by applying the magnetic field in opposite direction (- 5mT).} It means that we can also control the degeneracy of ASI structures by maneuvering the external field. Interestingly, the magnetization state of the individual peanut-shaped nanomagnet changes coherently, which implies that the horizontally aligned peanut-nanomagnets change their magnetization together. Likewise, the vertically arranged peanut-nanomagnets also change their magnetization state in a perfectly coherent way.  It can be attributed to the dipolar interaction as the corresponding demagnetization energy is always larger than the exchange energy counterpart as evident from Fig.~\ref{figure5}(b), which shows that the strength of the demagnetization energy is about one order higher in comparison to the exchange energy.

%{\color{black} The isolated vertices in both cases show two-in two-out (type-II) state in remanence corresponding to one of the sixteen possible magnetic degrees of freedom for such configuration as shown in Fig. 1.}

%{\color{black} In typical experiments where arrays of such vertices are studied, the edges of the vertices are practically closed with similar nanomagnets. Therefore, in order to understand the role of edge-islands in the micro-magnetic behavior of the vertex, we include eight nanomagnets of identical dimensions to close the edges of the isolated vertex.}

\begin{figure}[!htb]
	\centering\includegraphics[scale=0.10]{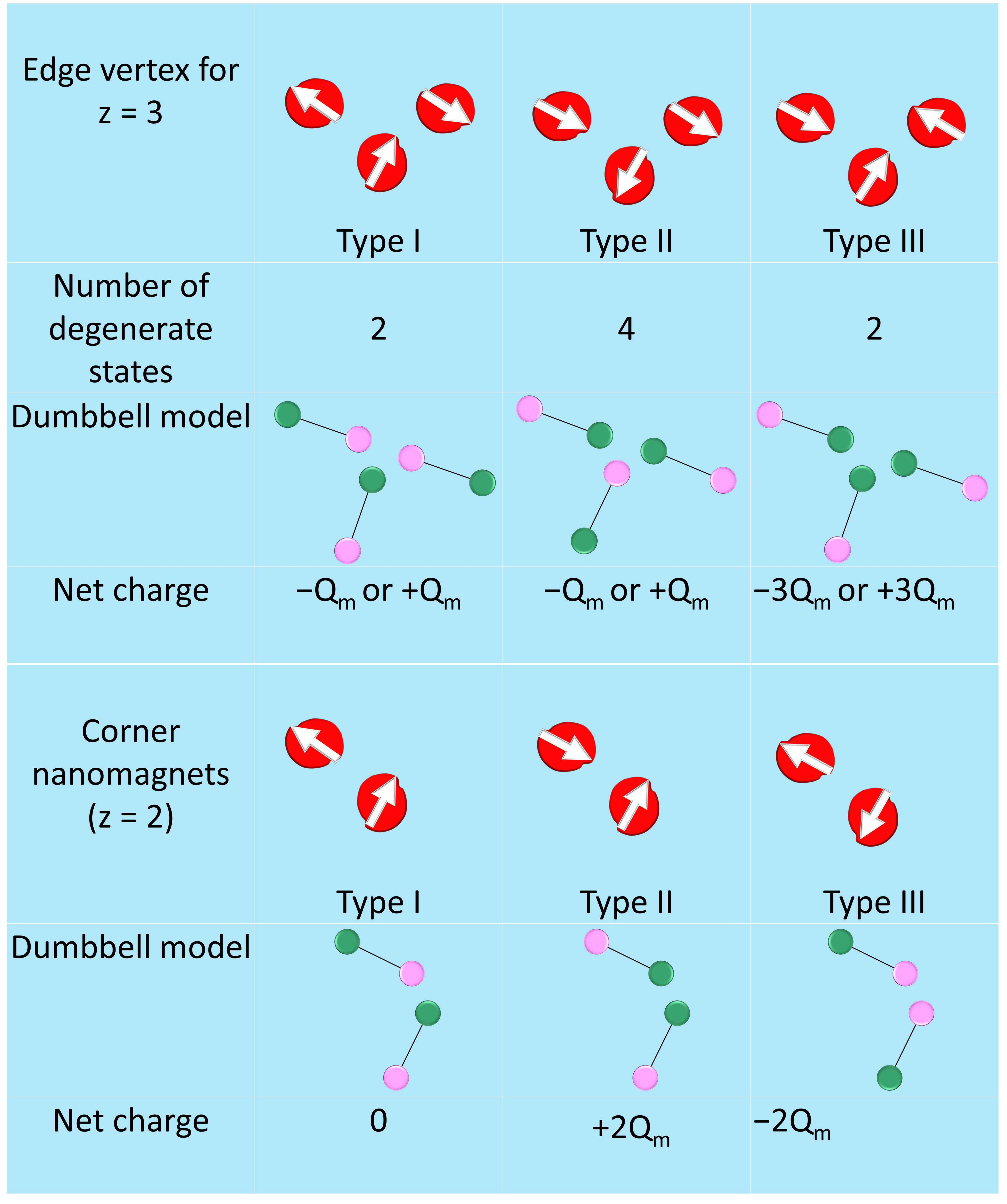}
	\caption{Schematics of the spin states at the edges with $z=3$ and $z=2$. Here, $z$ is defined as the total  number of the nanomagnets involved at the vertex edge and corner of ASI system. The corresponding degeneracy and the charge at the vertex edges for all possible state are shown as defined by dumbbell model.}
 \label{figure2}
\end{figure}

\begin{figure}[!htb]
	\centering\includegraphics[scale=0.10]{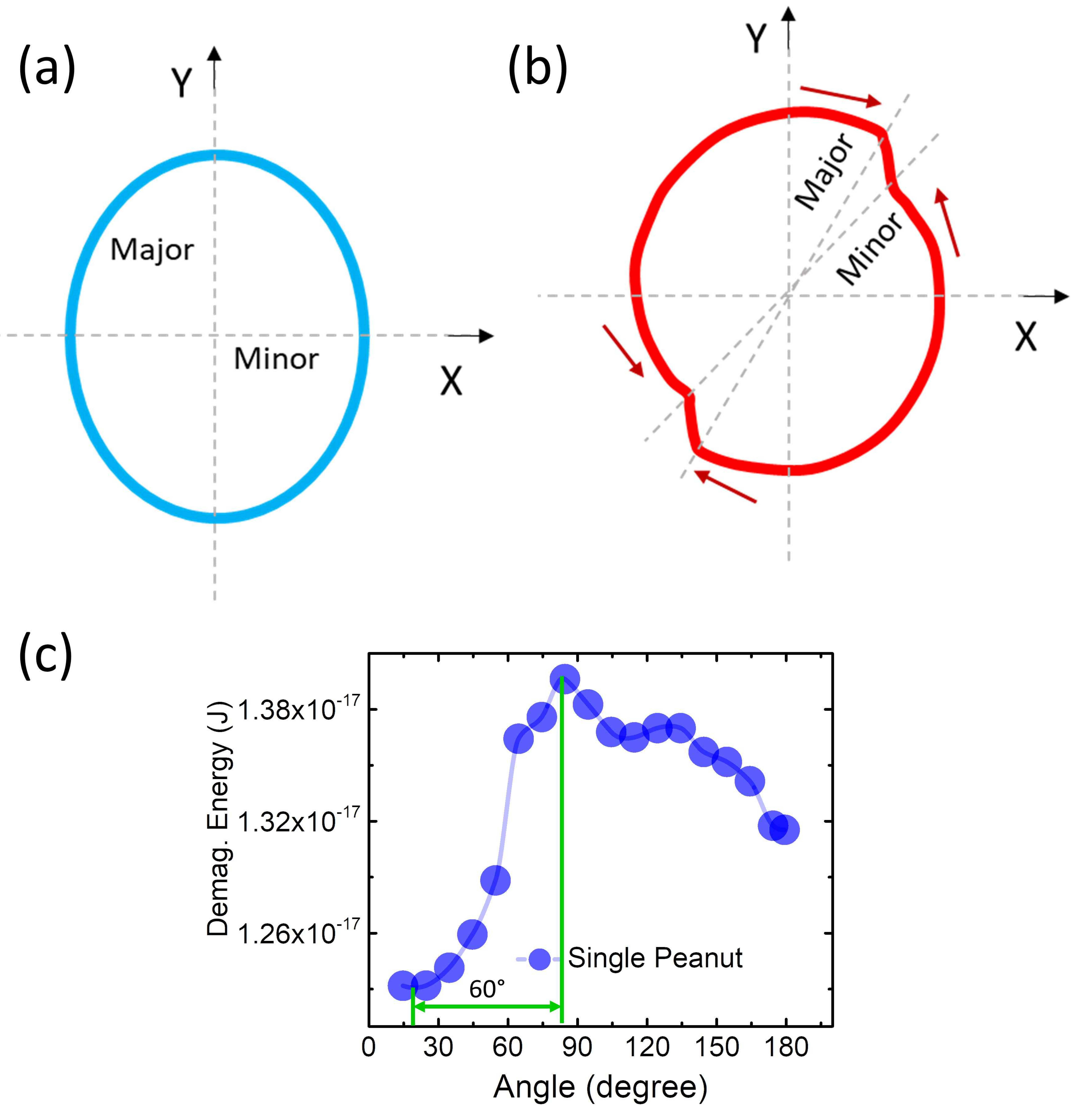}
	\caption{Schematic of {\color{black} hard} and {\color{black} easy} axis in the (a) elliptical and (b) peanut-shaped nanomagnet. (c) Variation of demagnetization energy with the rotation of applied magnetic field for peanut-shaped nanomagnet.}
	\label{figure3}
\end{figure}

\begin{figure}[!htb]
	\centering\includegraphics[scale=0.10]{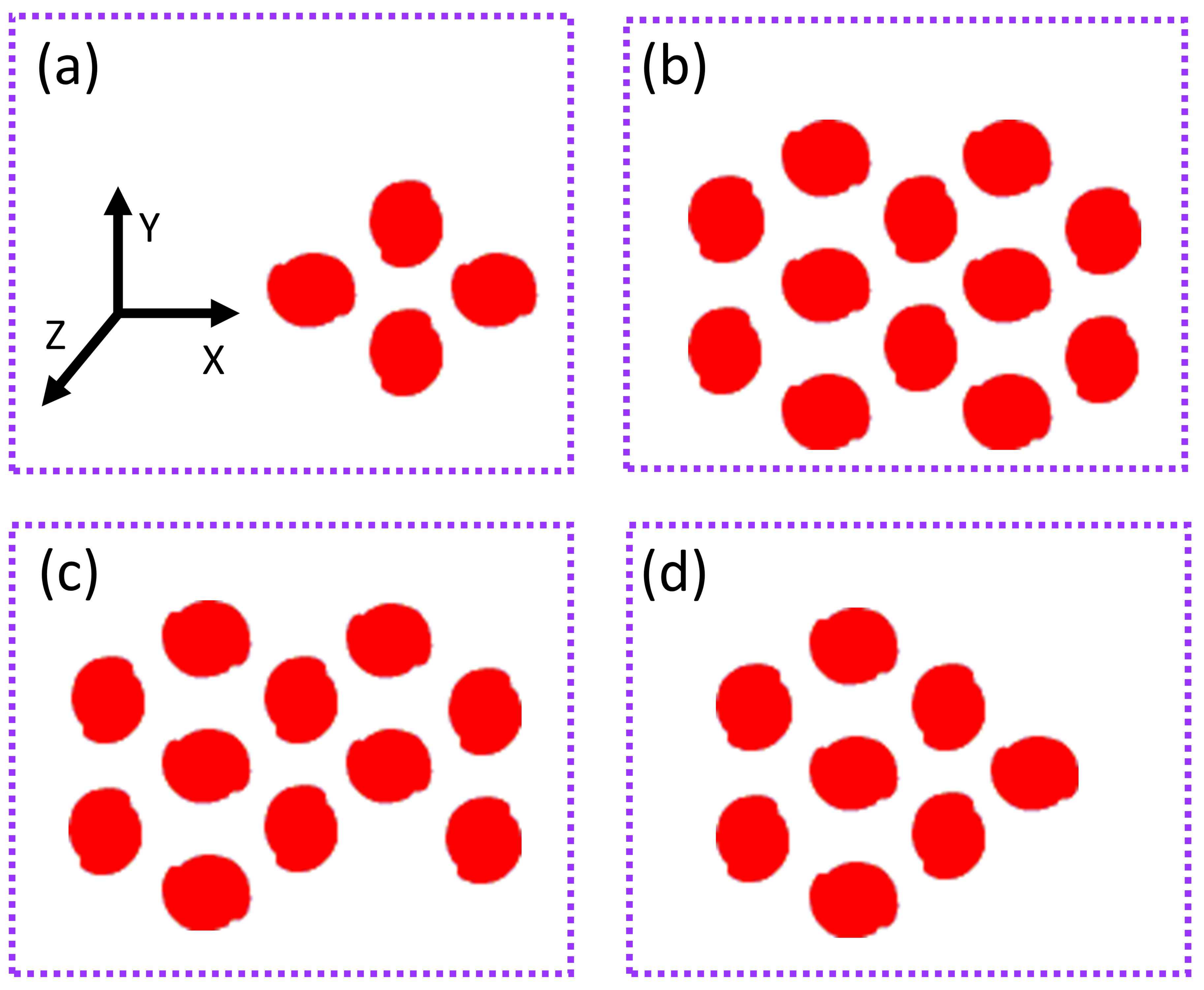}
	\caption{Schematics of the 2-dimensional peanut shape ASI system (a) vertex with open edges (b) vertex with closed edges (c-d) vertex with defect edges.}
	\label{figure4}
\end{figure}

\begin{figure}[!htb]
	\centering\includegraphics[scale=0.100]{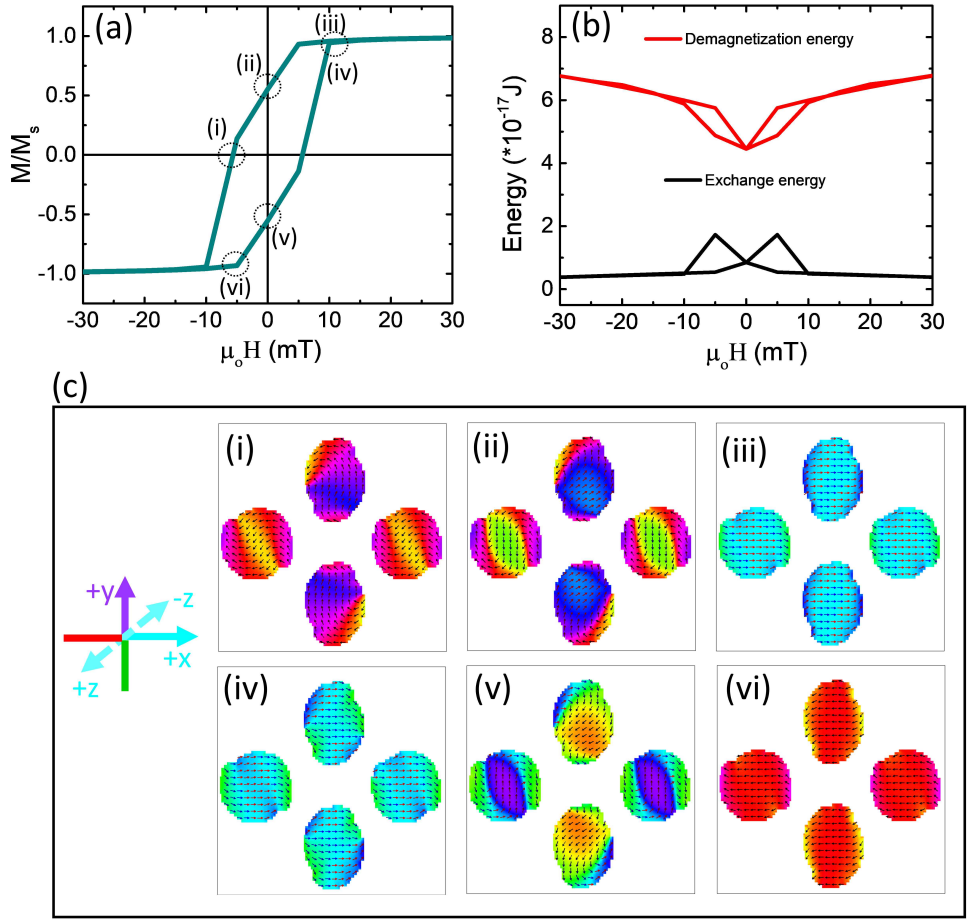}
	\caption{(a) Variation of the magnetization with the magnetic field for a vertex with open edges and (b) Variation of exchange and demagnetization energy with the magnetic field. (c) The magnetization configurations at various representative fields are shown. Type II state is clearly evident at remanence.}
	\label{figure5}
\end{figure}

\begin{figure}[!htb]
	\centering\includegraphics[scale=0.10]{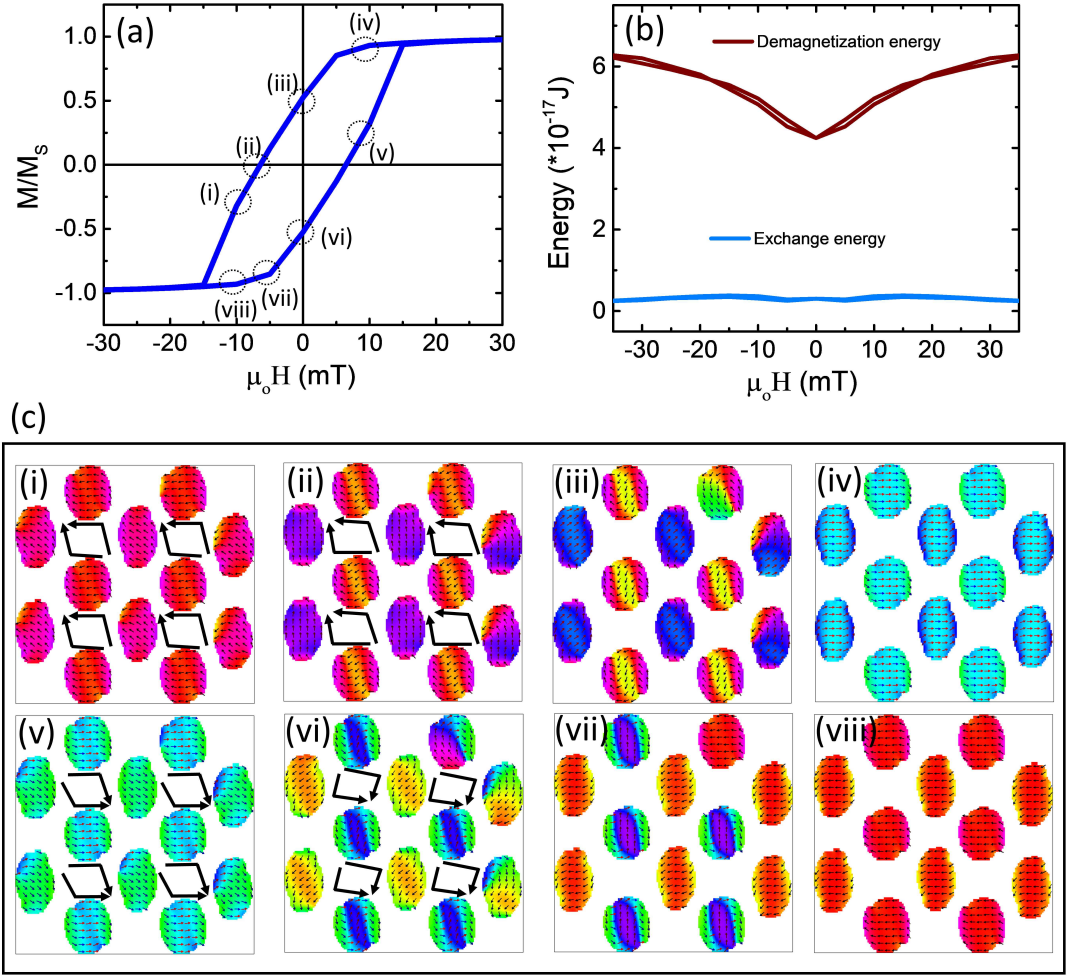}
	\caption{(a) Hysteresis for magnetization switching states of a vertex with closed edges and (b) Variation of exchange and demagnetization energy with the magnetic field.  (c) The magnetization configurations at various representative fields are shown. At remanance, the Type II state is clearly seen. The other possible microstates corresponding to Type II state is also accessed by suitable value of external magnetic field.}
	\label{figure6}
\end{figure}

Usually, in studying the magnetic behaviour of an arrays of such vertices, the edges are practically surrounded by similar nanomagnets. Therefore, to understand the role of edge-islands on the micro-magnetic behaviour of the vertex, we now consider ASI with eight nanomagnets of identical dimensions to close the edges of the isolated vertex as depicted in Fig.~\ref{figure4}(b). We plot the magnetization as a function of the external magnetic field in Fig.~\ref{figure6}(a) for the closed edge vertex. 
To further probe the role of the demagnetization and the exchange energy in magnetization switching, we have plotted them as a function of an external magnetic field. 
%to the applied magnetic field and found that 
The magnetization states are also shown at various magnetic fields, as depicted in Fig.~\ref{figure6}(c) for the field position (i) to (viii) marked in Fig.~\ref{figure6}(a). The number of significant jumps in the field-dependent magnetization curve is eight; the magnetic configuration of each such switching is shown in Fig.~\ref{figure6}(c). The first magnetization switching occurs at -10 mT. There are a number of interesting things to note at this point [Fig.~\ref{figure6}c(i)]: (1) 2-in/2-out configuration emerges at remanence in this case also. (2) Four onion-type structures also emerge at the edges as depicted in Fig.~\ref{figure6}c(i). (3) The four vertices with $z = 3$ at the edges are in either 2-in/1-out or 2-out/1-in state and are naturally  magnetically charged with $+Q_m$ or $-Q_m$ [please see the schematic Fig.~(\ref{figure2}) for reference]. (4) The four corners with $z=2$ have an absolute magnetic charge of $2Q_m$ or zero [see Fig.~(\ref{figure2}) for clarity]. If we consider all the magnetic charge in the system, it comes out to be zero. It means that the system maintains magnetic charge neutrality even after magnetization of individual nanomagnets gets flipped due to the external magnetic field. This magnetization switching is dominated by dipolar interaction as the demagnetization energy corresponding it is $5.07\times10^{-17}$ J, which is one order magnitude larger than the exchange energy ($=3.07\times10^{-18}$ J) counterpart. The next jump occurs at coercive field $\mu^{}_oH^{}_c=-6$ mT, and 2-in/2-out (Type II) configuration persists at the central vertex as shown in Fig.~\ref{figure6}c(ii). The dipolar interaction also dominates in this case as the demagnetization energy is $4.69\times10^{-17}$ J while the exchange energy equals  $2.86\times10^{-18}$ J. Similar observations can be drawn for the magnetic state observed at 0 mT, as depicted in Fig.~\ref{figure6}c(iii). Interestingly, we are able to access one of the other possible configurations corresponding to the Type II magnetic state at magnetic field strength of 10 mT and 0 mT [see Fig.~\ref{figure6}c(v)-(vi)]. Four onion type configurations still exist, and the system maintains magnetic charge neutrality in these cases also. The accessing of various possible configurations for a given state [see Fig.~(\ref{figure1})] clearly indicate that we can control the degeneracy of the system by an application of suitable magnetic field. Remarkably, the vertically and horizontally aligned peanut nanomagnets behave as if they are locked. Consequently, the horizontally placed nanomagnets change their magnetization in unison. Likewise, vertically aligned nanomagnets change their magnetization together. It is because the demagnetization energy is always larger than the exchange energy (see Fig.~\ref{figure6}(b)).

%The coercive field and remanent magnetization are approximately same as that of Type a geometry. The rate of change of magnetization with applied magnetic field is bit slow as compared to Type a structure. This could be due to an increase in dipolar and exchange interaction. It is interesting to note that two vertically aligned peanut (three in total) changes their magnetization in unison as if their magnetizations get locked. The same thing can be observed for three vertically aligned peanut (two in total). This is due that fact that with increase in dipolar and exchnage interaction, ferromagnetic coupling gets enhanced. As a results of this we may have multiple degenrate state (Energy will be the same whenever magnetic moments orient themselves in a coherent manner).

\begin{figure}[!htb]
	\centering\includegraphics[scale=0.10]{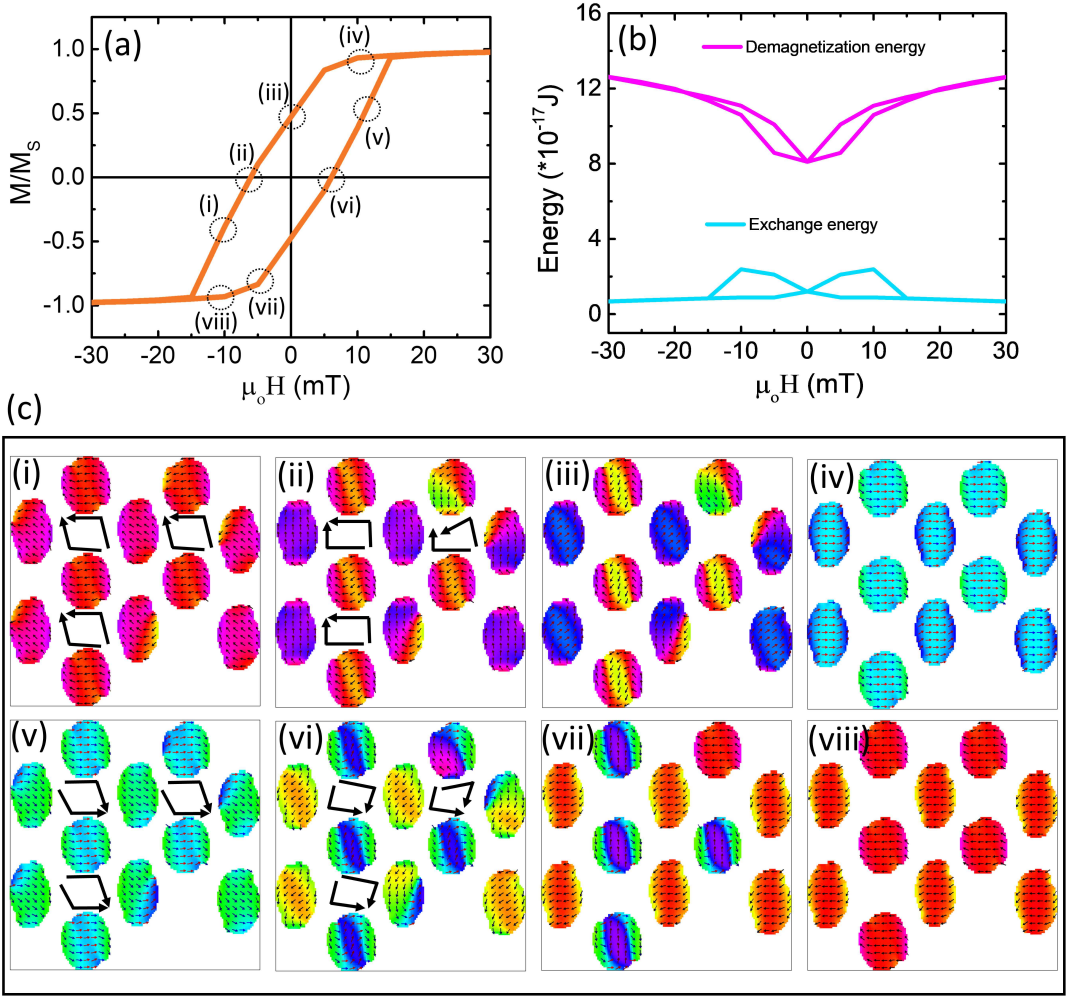}
	\caption{(a) Hysteresis for magnetization switching states of a vertex with defect edges and (b) Variation of exchange and demagnetization energy with the magnetic field (c) Corresponding distinct spin states at eight representative  magnetic field value. The defect is found to break the magnetic charge neutrality.}
	\label{figure7}
\end{figure}

The study of magnetization switching on the square-ASI system consisting of peanut-shaped nanomagnets with artificial defects in the lattice can give a better insight into the role of dipolar interaction on frustration in such an exotic system. These studies can also help analyze the frustration in pyrochlore spin ice systems because of controlled site vacancies. Therefore, we have performed systematic studies for such lattice defect-induced magnetic states in the square-ASI system for multiple defect configurations. Here, we report the results for two such defect configurations. In Fig.~(\ref{figure7}), we show the magnetization variation as a function of an external magnetic field of an isolated vertex with a lattice defect in the form of a missing nanomagnet at the edge of the two-dimensional square ASI structure [see schematic Fig.~\ref{figure4}(c) for clarity]. The magnetic configurations are also shown at various magnetic field value [see Fig.~\ref{figure7}(c)]. The total number of notable jumps in the magnetization is eight, indicating the occurrence of magnetization switching. The first switching occurs at -10 mT [see Fig.~\ref{figure7}c(i)]. Remarkably, the vertex retains the 2-in/2-out Type-II magnetic state at this magnetic field. The next magnetization switching occurs at $\mu^{}_oH^{}_c=-6$ mT. The central vertex retains the Type II magnetic state in this case also. In these two cases, there are also other important properties to note viz. (1) The magnetization switching is dominated by the dipolar interaction as the demagnetization energy $\approx10^{-16}$ J is one order larger than the exchange energy ($\approx10^{-17}$ J) as evident from the field dependent demagnetization and exchange energy curve (see Fig.~\ref{figure7}(b)).
 Interestingly, these states have also three distinct onion configurations, as depicted in Fig.~\ref{figure7}c(i)-(ii). (3) The three vertices with $z = 3$ at the edges are in either 2-in/1-out or 2-out/1-in state and are naturally charged with $+Q_m$ or $-Q_m$, similar to that of a defect-free situation.(4) The three corners with $z=2$ also have an absolute magnetic charge of $2Q_m$ or zero as that of a defectless system. But if we sum all the magnetic charges in the system, it comes out to be $3Q_m$, which is non-zero. It means that the defect (missing nanomagnet) breaks the magnetic charge neutrality of the underlying system. Therefore, the system behaves as an emergent magnetic monopole as a whole. We have also been able to access another possible configuration corresponding to Type II magnetic state at 10 and 0 mT [see Fig.~\ref{figure7}c(v)-(vi)]. These states also consist of three onion type configurations. Remarkably, such magnetization switching is also promoted by the dipolar interaction as the corresponding demagnetization energy ($\approx 1.10\times10^{-16}$ J) is larger than the exchange energy ($\approx 8.86\times10^{-18}$ J) as shown in Fig.~\ref{figure7}(b) . It is interesting to note that vertically aligned peanuts changes their magnetization orientation in unison except near the defect, which could be due to symmetry breaking induced by a defect.

\begin{figure}[!htb]
	\centering\includegraphics[scale=0.10]{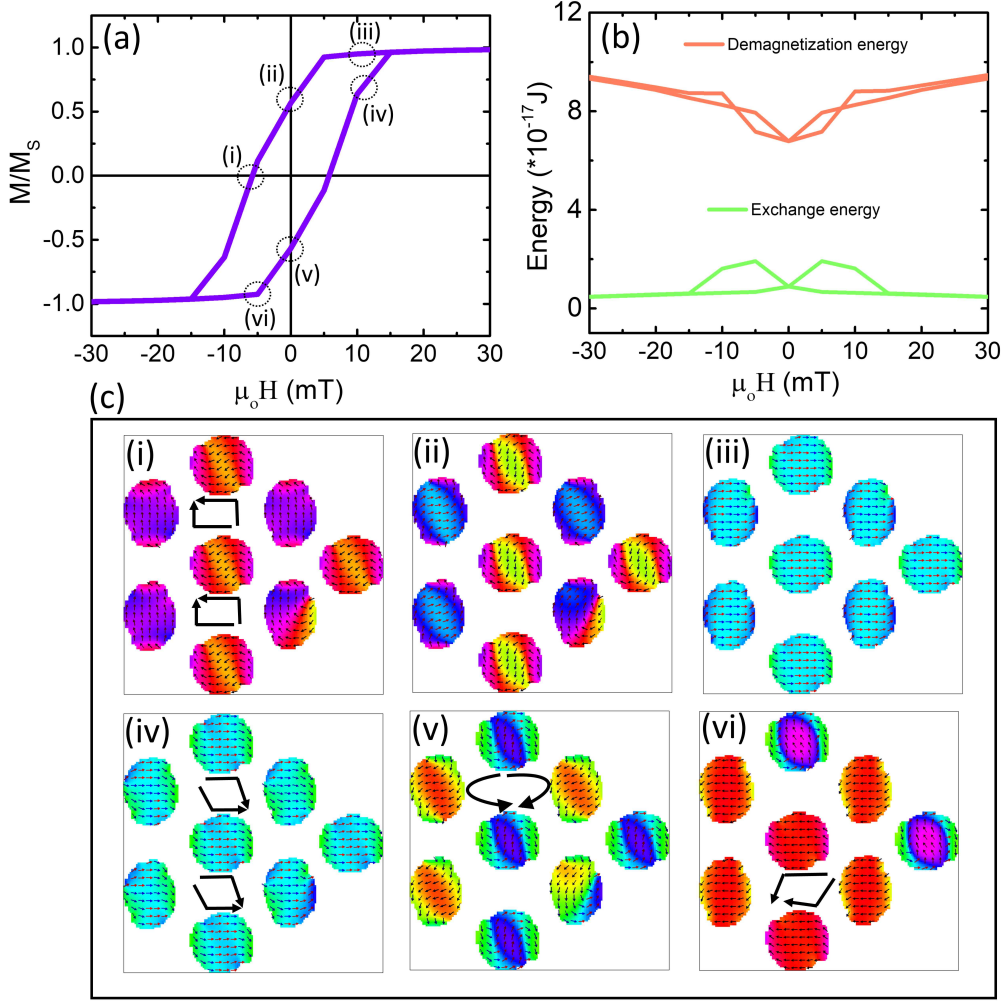}
	\caption{(a) Hysteresis for magnetization switching states of a vertex with defect edges and (b) Variation of exchange and demagnetization energy with the magnetic field (c) The magnetization states corresponding to six values of magnetic field. The magnetic charge neutrality is found to break in this case also. Interestingly, there is an emergence of Type III spin state [Fig.~\ref{figure8}c(iv)], indication of magnetic monopole.}
	\label{figure8}
\end{figure}

Finally, we study the micromagnetic behaviour of the ASI system with the enhanced defects. In Fig.~(\ref{figure8}), we plot the magnetic field dependence of magnetization and magnetization state for an ASI structure with four nanomagnets removed [please see the schematic Fig.~\ref{figure4}(d)]. The first switching occurs at $\mu_oH^{}_c\approx -6$ mT. The system retains a 2-in/2-out (Type II) configuration at this magnetic field, as depicted in Fig.~\ref{figure8}c(i). This state has two onion type configurations, and the system's magnetic charge-neutrality is found to break due to the defects. Therefore, the underlying system behaves as magnetic monopole. At a field value of 10 mT, the vertex appears to have a 3-in/1-out configurations, which is Type III microstate [see Fig.~\ref{figure8}c(iv)]. So the defects break the spin-ice rule, and a magnetic monopole emerges. The other possible configurations corresponding to Type II magnetic state emerges at 0 mT, which clearly indicates that we can control the degeneracy of the magnetization state. Remarkably, the vertex appears to have 1-in/3-out (Type III) magnetic state at -5 mT, which is also an indication of the emergence of a magnetic monopole. Because of the symmetry breaking, the dipolar field is not able to change the magnetization of the constituents nanomagnets in unison.

%Effect of controlled site vacancies on the frustration in pyrochlore spin ice systems can be studied much easily in ASI system by creating artificial defects in the form of missing nanomagnets.

%Systematic studies were performed for such lattice defect-induced magnetic states in the square-ASI system for mutliple defect configurations. 
 
%Here, we report the results of two such cases. Fig. 3(a) shows the remanent magnetic state of an isolated vertex with a lattice defect in the form of a missing nanomagnet at the edge of the two-dimensional square ASI structure. The vertex retains the two-in two-out type-II magnetic state at the remanence.

%We create defect in the structure by removing one of the peanut as shown in Fig.~(\ref{figure4}). In this case, the corecive field is slightly higher as compared to the two former structures (Type a and Type b). It is interesting to note that vertically aligned peanuts changes their magnetization orientation in unnison except near the defect. It means that degenerecy gets lifted in this case.  

\section{summary and conclusion}
%Now, we summarize and discuss the main results presented in this work.
In summary, we have systematically analyzed the micromagnetic states and magnetization switching behaviour of the proposed ASI system consisting of peanut-shaped nanomagnets arranged in a square geometry. The switching field is smaller by one order in comparison to conventionally used highly-anisotropic nanomagnets such as elliptical-shaped nanoisland. 
%It clearly strengthens the fact that we can control the magnetization of the constituent nanomagnets using piezoelectric strain easily in the present system, which has significant implications in energy-efficient spintronics based applications. 
Our extensive simulations show that the Type II magnetic state (2-in/2-out) is robust at the remanence. We also have accessed other possible microstates corresponding to Type II configurations by applying a suitable value of the magnetic field. It implies that  the degeneracy of the magnetic state can be controlled by manipulating the amplitude and direction of an external magnetic field. %This control of degeneracy is of vital importance in controlling the residual entropy in such an exotic spin ice system.

The interplay of dipolar interaction and lattice defects in the form of missing components of the ASI system  significantly affects the magnetization switching behaviour. They are also found to break the magnetic charge neutrality. As a result, the system behaves like a magnetic monopole as a whole. Our results also show that one can access other possible microstates by carefully maneuvering the defects and the external magnetic field. Our results suggest the long-ranged nature of dipolar interaction and the introduction of lattice defects can be exploited to predictably form a vertex of non-zero magnetic charge of Type-III spin ice state, reminiscent of an emergent magnetic monopole. These studies are of immense importance in analyzing dipolar interaction's role on frustration in a more profound way. {\color{black} Although the present work dealt with two-dimensional ASI system, but we believe that some of essential features of frustrations and other related physics of the system with higher dimension can be understood using the present work. Therefore, the study of switching mechanism of ASI system with defects in the form of controlled vacancies (missing nanomagnet) could be helpful in analyzing the frustration in pyrochlore spin ice systems}. It will be important to study the possibility of controlling the motion of such defect-induced magnetic states present in arrays. Therefore, we hope that our work opens up a platform for combined efforts in experimental, analytical, and computational studies for these useful and versatile systems.

\section*{Acknowledgment}
AC would like to acknowledge the NTU research scholarship (NTU-RSS) for funding support. R.S.R. would like to acknowledge the Ministry of Education (MOE), Singapore through Grant No. MOE2019-T2-1-058 and National Research Foundation (NRF) through Grant No. NRF-CRP21-2018-003.

\bibliographystyle{h-physrev}
\bibliography{refs}

\begin{thebibliography}{10}

\bibitem{nisoli2010}
C.~Nisoli, J.~Li, X.~Ke, D.~Garand, P.~Schiffer, and V.~H. Crespi,
\newblock Physical review letters {\bf 105}, 047205 (2010).

\bibitem{lendinez2019}
S.~Lendinez and M.~Jungfleisch,
\newblock Journal of Physics: Condensed Matter {\bf 32}, 013001 (2019).

\bibitem{skjaervo2020}
S.~H. Skj{\ae}rv{\o}, C.~H. Marrows, R.~L. Stamps, and L.~J. Heyderman,
\newblock Nature Reviews Physics {\bf 2}, 13 (2020).

\bibitem{kapaklis2014}
V.~Kapaklis, U.~B. Arnalds, A.~Farhan, R.~V. Chopdekar, A.~Balan, A.~Scholl,
  L.~J. Heyderman, and B.~Hj{\"o}rvarsson,
\newblock Nature nanotechnology {\bf 9}, 514 (2014).

\bibitem{nisoli2013}
C.~Nisoli, R.~Moessner, and P.~Schiffer,
\newblock Reviews of Modern Physics {\bf 85}, 1473 (2013).

\bibitem{zhang2019}
X.~Zhang, Y.~Lao, J.~Sklenar, N.~S. Bingham, J.~T. Batley, J.~D. Watts,
  C.~Nisoli, C.~Leighton, and P.~Schiffer,
\newblock APL Materials {\bf 7}, 111112 (2019).

\bibitem{wang2006}
R.~Wang, C.~Nisoli, R.~Freitas, J.~Li, W.~McConville, B.~Cooley, M.~Lund,
  N.~Samarth, C.~Leighton, V.~Crespi, {\em et~al.},
\newblock Nature {\bf 439}, 303 (2006).

\bibitem{morgan2011}
J.~P. Morgan, A.~Stein, S.~Langridge, and C.~H. Marrows,
\newblock Nature Physics {\bf 7}, 75 (2011).

\bibitem{jungfleisch2017}
M.~B. Jungfleisch, J.~Sklenar, J.~Ding, J.~Park, J.~E. Pearson, V.~Novosad,
  P.~Schiffer, and A.~Hoffmann,
\newblock Physical Review Applied {\bf 8}, 064026 (2017).

\bibitem{gilbert2014}
I.~Gilbert, G.-W. Chern, S.~Zhang, L.~O’Brien, B.~Fore, C.~Nisoli, and
  P.~Schiffer,
\newblock Nature Physics {\bf 10}, 670 (2014).

\bibitem{sen2015}
A.~Sen and R.~Moessner,
\newblock Physical review letters {\bf 114}, 247207 (2015).

\bibitem{iacocca2016}
E.~Iacocca, S.~Gliga, R.~L. Stamps, and O.~Heinonen,
\newblock Physical Review B {\bf 93}, 134420 (2016).

\bibitem{mamica2012}
S.~Mamica, M.~Krawczyk, and J.~W. K{\l}os,
\newblock Advances in Condensed Matter Physics {\bf 2012} (2012).

\bibitem{nikitin2015}
A.~A. Nikitin, A.~B. Ustinov, A.~A. Semenov, A.~V. Chumak, A.~A. Serga, V.~I.
  Vasyuchka, E.~L{\"a}hderanta, B.~A. Kalinikos, and B.~Hillebrands,
\newblock Applied Physics Letters {\bf 106}, 102405 (2015).

\bibitem{arava2019}
H.~Arava, N.~Leo, D.~Schildknecht, J.~Cui, J.~Vijayakumar, P.~M. Derlet,
  A.~Kleibert, and L.~J. Heyderman,
\newblock Physical Review Applied {\bf 11}, 054086 (2019).

\bibitem{wang2016}
Y.-L. Wang, Z.-L. Xiao, A.~Snezhko, J.~Xu, L.~E. Ocola, R.~Divan, J.~E.
  Pearson, G.~W. Crabtree, and W.-K. Kwok,
\newblock Science {\bf 352}, 962 (2016).

\bibitem{kapaklis2012}
V.~Kapaklis, U.~B. Arnalds, A.~Harman-Clarke, E.~T. Papaioannou, M.~Karimipour,
  P.~Korelis, A.~Taroni, P.~C. Holdsworth, S.~T. Bramwell, and
  B.~Hj{\"o}rvarsson,
\newblock New Journal of Physics {\bf 14}, 035009 (2012).

\bibitem{shen2012}
Y.~Shen, O.~Petrova, P.~Mellado, S.~Daunheimer, J.~Cumings, and
  O.~Tchernyshyov,
\newblock New Journal of Physics {\bf 14}, 035022 (2012).

\bibitem{qi2008}
Y.~Qi, T.~Brintlinger, and J.~Cumings,
\newblock Physical Review B {\bf 77}, 094418 (2008).

\bibitem{ladak2010}
S.~Ladak, D.~Read, G.~Perkins, L.~Cohen, and W.~Branford,
\newblock Nature Physics {\bf 6}, 359 (2010).

\bibitem{nascimento2012}
F.~Nascimento, L.~M{\'o}l, W.~Moura-Melo, and A.~Pereira,
\newblock New Journal of Physics {\bf 14}, 115019 (2012).

\bibitem{hugli2012}
R.~H{\"u}gli, G.~Duff, B.~O'Conchuir, E.~Mengotti, A.~F. Rodr{\'\i}guez,
  F.~Nolting, L.~Heyderman, and H.~Braun,
\newblock Philosophical Transactions of the Royal Society A: Mathematical,
  Physical and Engineering Sciences {\bf 370}, 5767 (2012).

\bibitem{castelnovo2008}
C.~Castelnovo, R.~Moessner, and S.~L. Sondhi,
\newblock Nature {\bf 451}, 42 (2008).

\bibitem{arava2020}
H.~Arava, E.~Y. Vedmedenko, J.~Cui, J.~Vijayakumar, A.~Kleibert, and L.~J.
  Heyderman,
\newblock Physical Review B {\bf 102}, 144413 (2020).

\bibitem{mengotti2011}
E.~Mengotti, L.~J. Heyderman, A.~F. Rodr{\'\i}guez, F.~Nolting, R.~V.
  H{\"u}gli, and H.-B. Braun,
\newblock Nature Physics {\bf 7}, 68 (2011).

\bibitem{loreto2014}
R.~Loreto, L.~Morais, R.~Silva, F.~Nascimento, C.~de~Araujo, L.~M{\'o}l,
  W.~Moura-Melo, and A.~Pereira,
\newblock arXiv preprint arXiv:1404.4082  (2014).

\bibitem{bramwell2012}
S.~T. Bramwell,
\newblock Philosophical Transactions of the Royal Society A: Mathematical,
  Physical and Engineering Sciences {\bf 370}, 5738 (2012).

\bibitem{ladak2011}
S.~Ladak, D.~Read, W.~Branford, and L.~Cohen,
\newblock New Journal of Physics {\bf 13}, 063032 (2011).

\bibitem{keswani2018}
N.~Keswani and P.~Das,
\newblock AIP Advances {\bf 8}, 101501 (2018).

\bibitem{chavez2018}
A.~C. Chavez, A.~Barra, and G.~P. Carman,
\newblock Journal of Physics D: Applied Physics {\bf 51}, 234001 (2018).

\bibitem{kaffash2021}
M.~T. Kaffash, S.~Lendinez, and M.~B. Jungfleisch,
\newblock Physics Letters A , 127364 (2021).

\bibitem{frotanpour2020}
A.~Frotanpour, J.~Woods, B.~Farmer, A.~P. Kaphle, E.~Lance, L.~Giovannini, and
  F.~Montoncello,
\newblock Physical Review B {\bf 102}, 224435 (2020).

\bibitem{chaurasiya2020nickel}
A.~Chaurasiya, P.~Pal, J.~V. Vas, D.~Kumar, S.~Piramanayagam, A.~Singh,
  R.~Medwal, and R.~Rawat,
\newblock Ceramics International {\bf 46}, 25873 (2020).

\bibitem{wolf2001}
S.~Wolf, D.~Awschalom, R.~Buhrman, J.~Daughton, v.~S. von Moln{\'a}r,
  M.~Roukes, A.~Y. Chtchelkanova, and D.~Treger,
\newblock science {\bf 294}, 1488 (2001).

\bibitem{hu2009}
J.-M. Hu and C.~Nan,
\newblock Physical Review B {\bf 80}, 224416 (2009).

\bibitem{wang2014}
J.~Wang, J.~Hu, J.~Ma, J.~Zhang, L.~Chen, and C.~Nan,
\newblock Scientific reports {\bf 4}, 1 (2014).

\bibitem{keswani2020}
N.~Keswani, R.~Lopes, Y.~Nakajima, R.~Singh, N.~Chauhan, T.~Som, S.~Kumar,
  A.~Pereira, and P.~Das,
\newblock arXiv preprint arXiv:2011.06860  (2020).

\bibitem{keswani2019}
N.~Keswani and P.~Das,
\newblock Journal of Applied Physics {\bf 126}, 214304 (2019).

\bibitem{buzzi2013}
M.~Buzzi, R.~V. Chopdekar, J.~L. Hockel, A.~Bur, T.~Wu, N.~Pilet, P.~Warnicke,
  G.~P. Carman, L.~J. Heyderman, and F.~Nolting,
\newblock Physical review letters {\bf 111}, 027204 (2013).

\bibitem{cui2017}
J.~Cui, S.~M. Keller, C.-Y. Liang, G.~P. Carman, and C.~S. Lynch,
\newblock Nanotechnology {\bf 28}, 08LT01 (2017).

\bibitem{salehi2017}
M.~Salehi-Fashami and N.~D’Souza,
\newblock Journal of Magnetism and Magnetic Materials {\bf 438}, 76 (2017).

\bibitem{wang2014full}
J.~Wang, J.~Hu, J.~Ma, J.~Zhang, L.~Chen, and C.~Nan,
\newblock Scientific reports {\bf 4}, 1 (2014).

\bibitem{anand2016}
M.~Anand, J.~Carrey, and V.~Banerjee,
\newblock Physical Review B {\bf 94}, 094425 (2016).

\bibitem{miltat2007}
J.~E. Miltat and M.~J. Donahue,
\newblock Handbook of Magnetism and Advanced Magnetic Materials  (2007).

\bibitem{arora2021}
N.~Arora and P.~Das,
\newblock AIP Advances {\bf 11}, 035030 (2021).

\bibitem{donahue1999}
M.~Donahue and D.~Porter,
\newblock National Institute of Standards and Technology, Gaithersburg, MD {\bf
  29}, 53 (1999).

\bibitem{gartside2018}
J.~C. Gartside, D.~M. Arroo, D.~M. Burn, V.~L. Bemmer, A.~Moskalenko, L.~F.
  Cohen, and W.~R. Branford,
\newblock Nature nanotechnology {\bf 13}, 53 (2018).

\bibitem{carrey2011}
J.~Carrey, B.~Mehdaoui, and M.~Respaud,
\newblock Journal of Applied Physics {\bf 109}, 083921 (2011).

\bibitem{remhof2008}
A.~Remhof, A.~Schumann, A.~Westphalen, H.~Zabel, N.~Mikuszeit, E.~Vedmedenko,
  T.~Last, and U.~Kunze,
\newblock Physical Review B {\bf 77}, 134409 (2008).

\end{thebibliography}
\newpage
\newpage
\newpage
\newpage

\newpage

\newpage

\newpage
\newpage

\end{document}